\documentclass[12pt]{article}

\usepackage[T1]{fontenc} 

\usepackage{microtype}

\usepackage[margin=1in]{geometry}
\usepackage[english]{babel}
\usepackage[autostyle, english=american]{csquotes}
\MakeOuterQuote{"}

\usepackage{amsmath,amssymb,amsfonts,amsthm}
\usepackage{xfrac,bm,upgreek}

\usepackage{graphicx}
\graphicspath{{figures/}}

\usepackage{booktabs,multirow,tabularx,dcolumn,threeparttable,makecell,array,longtable}
\newcolumntype{R}{>{\raggedright\arraybackslash}p{6cm}}
\newcolumntype{Q}{>{\raggedright\arraybackslash}p{6cm}}

\usepackage{caption}
\usepackage{subcaption}      

\usepackage{float}
\usepackage{pdflscape}
\usepackage{rotating}
\usepackage{afterpage}

\usepackage{enumitem}
\usepackage{ragged2e}

\usepackage{setspace}

\usepackage{eurosym}

\usepackage{xcolor}
\definecolor{medium-blue}{rgb}{0,0,0.5}
\definecolor{ChadBlue}{rgb}{.1,.1,.5}
\definecolor{ChadRed}{rgb}{.5,0,.5}

\usepackage{xurl}

\usepackage[
  colorlinks=true,
  urlcolor=medium-blue,
  linkcolor=ChadRed,
  citecolor=ChadBlue
]{hyperref}

\usepackage[all]{hypcap}

\usepackage{natbib}
\usepackage{sectsty}
\sectionfont{\large}

\usepackage[title]{appendix}

\usepackage{xr}
\makeatletter
\newcommand*{\addFileDependency}[1]{%
  \@addtofilelist{#1}%
  \IfFileExists{#1}{}{\typeout{No file #1.}}%
}
\newcommand*{\myexternaldocument}[1]{%
  \IfFileExists{#1.aux}{%
    \externaldocument{#1}%
    \addFileDependency{#1.tex}%
    \addFileDependency{#1.aux}%
  }{}%
}
\makeatother
\myexternaldocument{}

\title{Reassessing the role of intermediaries in exports\thanks{We thank the Spanish Tax Agency's Department of Customs and Excise (AEAT) for customs data. We thank participants at the Deusto Business School Brown Bag Seminar and XXVIII Encuentro de Economía Aplicada for their comments and suggestions. This research was conducted as part of Project PID2021-122133NB-I00 financed by MCIN/AEI/10.13039/501100011033/FEDER, EU. We also gratefully acknowledge financial support from the Basque Government Department of Education (IT1429-22 and IT1793-26).}}

\author{%
\large Aitor Garmendia-Lazcano\thanks{Garmendia-Lazcano: Deusto Business School, University of Deusto, Camino de Mundaiz 50, 20012 Donostia--San Sebasti\'an (Spain). Email: \href{mailto:aitor.garmendia@deusto.es}{aitor.garmendia@deusto.es}}%
\and
\large Ra\'ul M\'inguez\thanks{M\'inguez: C\'amara de Comercio de Espa\~na and Universidad Antonio de Nebrija. Calle de Santa Cruz de Marcenado, 27, 28015 Madrid (Spain). Email: \href{mailto:rminguez@nebrija.es}{rminguez@nebrija.es}}%
\and
\large Asier Minondo\thanks{Minondo: Corresponding author. Deusto Business School, University of Deusto, Camino de Mundaiz 50, 20012 Donostia--San Sebasti\'an (Spain). Email: \href{mailto:aminondo@deusto.es}{aminondo@deusto.es}}%
}

\date{This version: \today}

\begin{document}
\maketitle

\begin{abstract}
Previous studies conclude that intermediaries account for a large share of exports. Using Spanish firm-level data, we show that many firms classified as intermediaries are either manufacturer-owned export arms that ship their parent firms' products or vertically integrated firms that control design, production, and distribution and predominantly export goods sold under their own brands. Once we exclude these export arms and vertically integrated firms, the share of intermediaries in exports in our sample falls by about 70\%. We also show that pure intermediaries differ markedly from export arms and vertically integrated firms along key firm and export dimensions.
\end{abstract}

\begin{flushleft}
\end{flushleft}

\newpage 
\onehalfspacing

\section{Introduction}
\label{sec:introduction}

How firms export is a central question in international trade. Early models largely emphasized a single route: firms manufacture a product and export it directly \citep{melitz2003impact}. With the advent of richer microdata, the literature has shown that many manufacturers do not export directly \citep{bernard2010wholesalers}. Instead, they sell to domestic intermediaries, which then handle exporting. Evidence from multiple countries indicates that intermediaries can represent more than 40\% of all exporters and account for up to 25\% of total goods exports.\footnote{See, among others, \cite{crozet2013wholesalersFrance}, \cite{bernard2015intermediaries}, \cite{akerman2018wholesalers}, or \cite{blum2018trade}.}

Export-intermediation models typically assume that intermediaries source products from manufacturers that are insufficiently productive or lack the trade capabilities needed to generate enough profit from foreign sales to cover the fixed costs of exporting \citep{ahn2011intermediaries,minguez2026hybrid}. In these models, intermediaries have no proprietary relationship with manufacturers; their sole function is to deploy superior trade capabilities to market and sell manufacturers' products abroad \citep{felbermayr2011intermediaries,crozet2013wholesalersFrance,fujii2017indirectexports,akerman2018wholesalers}. 

Empirical studies define export intermediaries as firms whose reported activity is wholesaling or retailing, or whose employees work in those activities \citep{bernard2010wholesalers,crozet2013wholesalersFrance}. Using Spanish data that allow us to identify wholesale and retail exporters on a firm-by-firm basis, this paper shows that many firms initially classified as intermediaries depart markedly from the export intermediary archetype. First, a sizable share of these firms are owned by producers and export exclusively their parent firms' products. In these cases, producers may establish distribution subsidiaries and export through them for reasons unrelated to a lack of productivity or trade capabilities, such as limiting legal exposure, organizing international operations, or centralizing tax and administrative functions. In other cases, the firm manages the distribution of a foreign company’s products in the Spanish market and also serves as an exporter of those products to other markets (e.g., Portugal). These firms therefore operate as producers' export arms, rather than as independent intermediaries of the type typically described in the literature.

Second, some firms classified as intermediaries oversee the entire value chain, from product design and manufacturing to distribution, and trade exclusively in goods marketed under their own brands. These vertically integrated brand retailers (common, for example, in the fashion industry) likewise differ sharply from the archetype emphasized in the literature, in which an intermediary markets products sourced from low-productivity manufacturers and plays no role in their design or production.

If exporters are classified as intermediaries solely on the basis of their self-reported wholesale or retail activity, they account for 40.2\% of all exporters and 26.0\% of exports in our 2024 sample. The main contribution of this paper is to show that, once we exclude export arms and vertically integrated firms, intermediaries' share of total exports falls by 70.3\%, from 26.0\% to 7.7\%. The number of firms classified as intermediary exporters also declines sharply, from 7,600 to 4,432, a 41.7\% reduction; as a share of all exporters, pure intermediaries fall from 40.2\% to 23.4\%.

We also show that pure intermediaries differ from export arms and vertically integrated firms along key firm and export dimensions. On the firm side, export arms and vertically integrated firms employ more workers and generate higher value added per employee than pure intermediaries. On the export side, they export larger values. These results suggest that grouping all wholesale and retail exporters into a single homogeneous category can lead to misleading conclusions about their characteristics and export behavior.

Our paper contributes to the literature quantifying the role of intermediaries in exports.\footnote{A related literature shows that some manufacturers export not only their own goods but also goods produced by other firms \citep{bernard2019carry,erbahar2023tradeintermediation}.} \citet{bernard2010wholesalers} show that, in the United States in 2002, wholesalers and retailers accounted for 43\% of exporters, but only 9\% of total exports. In Italy in 2003, wholesalers and retailers represented 35\% of exporters and 12\% of exports \citep{bernard2015intermediaries}. In Sweden in 2005, wholesalers accounted for 45\% of exporters and 14\% of exports \citep{akerman2018wholesalers}. For Chile in 2007, \citet{blum2018trade} report that wholesalers and retailers represented 34\% of exporters and 7\% of exports. For France in 2007, \citet{crozet2013wholesalersFrance} find that wholesalers accounted for 32\% of exporters and 20\% of exports. According to statistics published by the Spanish Tax Agency (Agencia Tributaria), wholesalers and retailers accounted for 41\% of all Spanish goods exporters and 25\% of the value of goods exports in 2023.\footnote{Agencia Tributaria, \emph{Datos de Comercio Exterior por características de la empresa}. Available at: \url{https://sede.agenciatributaria.gob.es}.} We contribute to this literature by showing that many wholesalers and retailers are not intermediaries. Instead, they are either export arms of producer firms or vertically integrated firms that export goods under their own brands. Note that classifying a wholesaler or a retailer as an intermediary solely because 100\% of its employment is in wholesaling or retailing, as in \cite{bernard2010wholesalers}, also has limitations. A firm may satisfy this criterion and still be an export arm or even a vertically integrated firm without in-house manufacturing activity. Excluding these categories, which depart markedly from the intermediary archetype used in trade models, substantially reduces intermediaries' shares in both exporters and exports.

The role of pure intermediaries in exports has important implications for export-promotion policy. If pure intermediaries account for a substantial share of exports, policies that facilitate matching between firms lacking the capabilities to serve foreign markets and specialized intermediaries may be effective in promoting exports. By contrast, if their contribution to aggregate exports is limited, policies aimed at strengthening trade-related capabilities within producing firms may yield larger export gains.

The remainder of the paper is organized as follows. The next section defines three categories of wholesale and retail exporters and describes the procedure used to assign each firm to a category. Section~\ref{sec:reassessing} reports the contribution of each category to the number of exporters and to total exports and presents descriptive regressions comparing the categories along key firm and export characteristics. The final section concludes.

\section{Identification and classification of export intermediaries}
\label{sec:categories_of_wholesalers}

Data on the universe of Spanish firms' export transactions in goods were obtained from the Customs and Excise Department of the Spanish Tax Agency (AEAT--Customs). Each export record reports the value, in euros, by firm, CN8 product, destination country, year, and month.\footnote{We exclude Harmonized System 2-digit codes 27 (mineral fuels) and 98--99 (special classifications) from the sample.}

AEAT--Customs does not report firms' main economic activity, so we merge the customs records with Bureau van Dijk's SABI database using the correspondence described in \cite{delucio2018prices}. SABI contains financial and accounting information for firms that file accounts with the Spanish Business Register and reports each firm’s 4-digit NACE Rev. 2 industry classification. Due to statistical confidentiality restrictions, we can use information only for firms included in the \textit{Directorio de Empresas Exportadoras e Importadoras}, a database compiled by the C\'amara de Comercio de Espa\~na (Spanish Chamber of Commerce). This restriction reduces the number of exporters in our sample by 74\% (from 72{,}120 to 18{,}883); however, these firms still account for 81\% of Spanish exports in 2024. That is, the restriction mainly removes very small exporters, which explains why the retained firms still account for most export value.

We classify a Spanish exporter as a wholesale or retail exporter if its 2-digit NACE Rev.\ 2 code is 46 or 47 (wholesale and retail activities, excluding sales of motor vehicles and motorcycles). We define an exporter as belonging to the primary sector if its NACE Rev.\ 2 code is between 01 and 09, and as a manufacturer if the code is between 10 and 32 (both included). The remaining exporters are classified as other activities.

Table~\ref{tab:table_exporters_exports_by_status} reports the number of exporters and the value of exports by main activity in our 2024 sample. Wholesale and retail exporters account for 40.2\% of exporters and 26.0\% of exports in our sample. Manufacturing is the largest category, comprising 48.6\% of exporters and 62.0\% of exports. Other activities represent 9.2\% of exporters and 10.8\% of exports, while the primary sector accounts for only a small fraction of both exporters and exports (2.0\% and 1.2\%, respectively).


\begin{table}[htbp]
\begin{center}
\caption{Exports by main activity, 2024}
\label{tab:table_exporters_exports_by_status}
\begin{tabular}{lrrrr}
\toprule
Activity & \# of exporters & \multicolumn{1}{c}{\shortstack{Share in total \#\\of exporters (\%)}} & \multicolumn{1}{c}{\shortstack{Exports\\(million euros)}} & \multicolumn{1}{c}{\shortstack{Share in total\\exports (\%)}}\\
\midrule
Primary &       373 & 2.0 &        3,239 & 1.2 \\
Manufacturing &     9,173 & 48.6 &      172,993 & 62.0 \\
Wholesale and retail & 7,600 & 40.2 &  72,518 & 26.0 \\
Other activities &     1,737 & 9.2 &       30,085 & 10.8 \\
\midrule
Total &    18,883 &     100.0 &      278,835 &        100.0 \\
\bottomrule
\end{tabular}
\caption*{\begin{footnotesize}Source: authors' calculations based on AEAT--Customs data merged with Bureau van Dijk's SABI database.\end{footnotesize}}
\end{center}
\end{table}

Our sample closely matches the full population of Spanish goods exporters in terms of the share of wholesale and retail firms in both the number of exporters and export value. Although the official statistics refer to 2023 and our sample refers to 2024, the shares are very similar: wholesalers and retailers account for 41\% of exporters in the full population and 40\% in our sample, and for 25\% of export value in the full population and 26\% in our sample.

We define three groups of wholesale and retail exporters. The first group comprises intermediaries. These firms correspond to the archetype of intermediary exporters: they purchase products from manufacturers with which they have no ownership ties and sell them in foreign markets. An example of an intermediary exporter is Quimidroga, S.A., a firm that distributes around 8,000 chemical product references.\footnote{\url{https://www.quimidroga.com/en/chemical-distributor/}.}

The second group is denoted as export arms and consists of distribution subsidiaries owned by primary-sector or manufacturing firms that mainly export their parent firms' products. An example of this type of firm is Natra Chocolate International, S.L. This company is the principal distribution and commercial subsidiary of the Natra Group, a Spanish multinational and global leader in the production of chocolate products and cocoa derivatives. It serves as the group's centralized platform for managing sales, distribution, and logistics across key European and international markets.\footnote{Data from SABI and \url{https://natra.com/es/}.} Many Spanish subsidiaries of foreign firms are also included in this group. These firms manage the distribution of a foreign company's products in the Spanish market and also export those products to other markets. An example is Adidas España, S.A.U., a company wholly owned by Adidas AG that distributes Adidas products in Spain and abroad.\footnote{Data from SABI and \url{https://www.adidas.es/}.}

The third group is composed of firms that coordinate the entire value chain, from product design and production to distribution, and export predominantly goods marketed under their own brands. An example of this type of firm is Punto Fa, S.L., the owner of the fashion brand Mango.\footnote{\url{https://shop.mango.com/es/es}.}

To assign each wholesale or retail exporter to the intermediary, export arm, or vertically integrated category, we rely on the reasoning-optimized versions of three large language models (LLMs): ChatGPT (version 5.4), Claude (Sonnet 4.6, Adaptive thinking), and Gemini (3.5 Thinking).\footnote{The classification exercise was conducted between April and June 2026.} The prompt used to query these LLMs, reproduced in Appendix~\ref{app_sec:prompt}, describes the defining characteristics of the three types of wholesale and retail exporters observed in practice and asks the models to classify each firm accordingly.

Table~\ref{tab:detailed_type_ia} in Appendix~\ref{app_sec:additional_tables} reports the classifications of wholesale and retail exporters generated by each LLM. In all three models, intermediaries constitute the largest category and vertically integrated firms account for a larger share of wholesale and retail exporters than export arms. At the same time, the distribution of firms across categories varies across LLMs. For example, the share of intermediaries is higher in ChatGPT (66.4\%) than in Claude (51.9\%) and Gemini (49.9\%), while the share of export arms is higher in Claude (19.3\%) than in ChatGPT (16.0\%) and Gemini (15.2\%).

Table~\ref{tab:detailed_llm_cross} in Appendix~\ref{app_sec:additional_tables} reports pairwise agreement across LLM classifications by wholesale and retail firm type. The table should be read by row. For example, in the comparison between ChatGPT and Claude, 69.1\% of the firms that ChatGPT classifies as intermediaries are also classified as intermediaries by Claude. Claude classifies the remaining 7.1\% of those firms as export arms and 23.8\% as vertically integrated firms. Overall agreement across classifications is around 70\%, with the highest level of agreement observed between ChatGPT and Gemini (73.0\%). Across firm types, the highest simple average matching rate is found for export arms (72.6\%), followed by intermediaries (71.0\%) and vertically integrated firms (68.7\%).

To obtain the final classification, we assign each wholesale or retail exporter to a category when at least two of the three LLMs place it in that category. For 4,286 of the 7,600 wholesale and retail exporters (56.4\%), the three LLMs produce the same classification. For another 3,181 firms (41.9\%), two of the three LLMs agree. This leaves 133 firms (1.7\%) for which each LLM assigns a different category; these cases are reviewed manually.

We also manually review the top 200 wholesale and retail exporters by export value, which account for 60.3\% of total wholesale and retail exports. In 97.0\% of these cases, our manual classification coincides with the final LLM-based classification. Finally, we manually review a random sample of 366 firms, corresponding to the sample size required for a 95\% confidence level and a 5-percentage-point margin of error under the conservative assumption of maximum variance. The manual classification and the LLM-based classification coincide in 90.7\% of these cases. In all cases of discrepancy, we use our manual classification.


\section{Results}
\label{sec:reassessing}

Table~\ref{tab:table_wholesalers} summarizes the distribution of wholesale and retail exporters after the classification procedure. Intermediaries are the largest category among wholesale and retail exporters, accounting for 23.4\% of all goods exporters in our sample and 58.3\% of wholesale and retail exporters. The second-largest group consists of vertically integrated firms, which represent 10.2\% of all exporters and 25.3\% of wholesale and retail exporters. Export arms account for 6.6\% of all exporters and make up 16.4\% of wholesale and retail exporters. Overall, focusing on the archetype emphasized in trade models (pure intermediaries), the reclassification yields a 41.7\% reduction in the number of intermediary exporters.

\begin{table}[hbtp]
\begin{center}
\caption{Wholesale and retail categories, 2024}
\label{tab:table_wholesalers}
\begin{tabular}{lrrrr}
\toprule
Category& \# of exporters& \multicolumn{1}{c}{\shortstack{Share in total \#\\of exporters (\%)}} & \multicolumn{1}{c}{\shortstack{Exports\\(million euros)}} & \multicolumn{1}{c}{\shortstack{Share in total\\exports (\%)}} \\
\midrule
Intermediary &     4,432 & 23.4 &       21,508 & 7.7 \\
Export arm &     1,242 & 6.6 &       20,800 & 7.5 \\
Vertically integrated &     1,926 & 10.2 &       30,210 & 10.8 \\
\midrule
Total &     7,600 &      40.2 &       72,518 &         26.0 \\
\bottomrule
\end{tabular}
\caption*{\begin{footnotesize}Source: authors' calculations based on AEAT--Customs data merged with Bureau van Dijk's SABI database and LLM classifications.\end{footnotesize}}
\end{center}
\end{table}

The composition of wholesale and retail exporters also changes markedly when firms are weighted by export value. Vertically integrated firms become the largest category, accounting for 10.8\% of total exports and 41.7\% of wholesale and retail exports. By a small margin, intermediaries are the second-largest group, representing 7.7\% of total exports and 29.7\% of wholesale and retail exports. Although they comprise only a small fraction of exporters, export arms account for a sizable 7.5\% of total exports and 28.6\% of wholesale and retail exports.

It is interesting to note that, although vertically integrated wholesalers and retailers represent a smaller share of exporters than intermediaries, they account for a larger share of total exports. This pattern is explained by the concentration of vertically integrated firms among the top wholesale and retail exporters. Figure~\ref{fig:ranking} ranks wholesale and retail exporters by export value, in descending order, using bins of 200 firms. The share of vertically integrated firms reaches its peak in the top-exporter bin (1-200), at 46\%, and declines as we move to lower-ranked bins. By contrast, the share of intermediaries rises as we move away from the top-exporter bin and reaches its highest levels among the lowest-value exporters. The share of export arms also declines gradually as we move away from the top-exporter bins.

\begin{figure}[t]
\label{fig:ranking}
\begin{center}
\begin{minipage}{\textwidth}
    \centering
    \caption{Distribution of wholesale and retail exporter types by export ranking (\%)}
    \label{fig:ranking}
    \hspace*{-0.9cm}\includegraphics[width=1.03\textwidth]{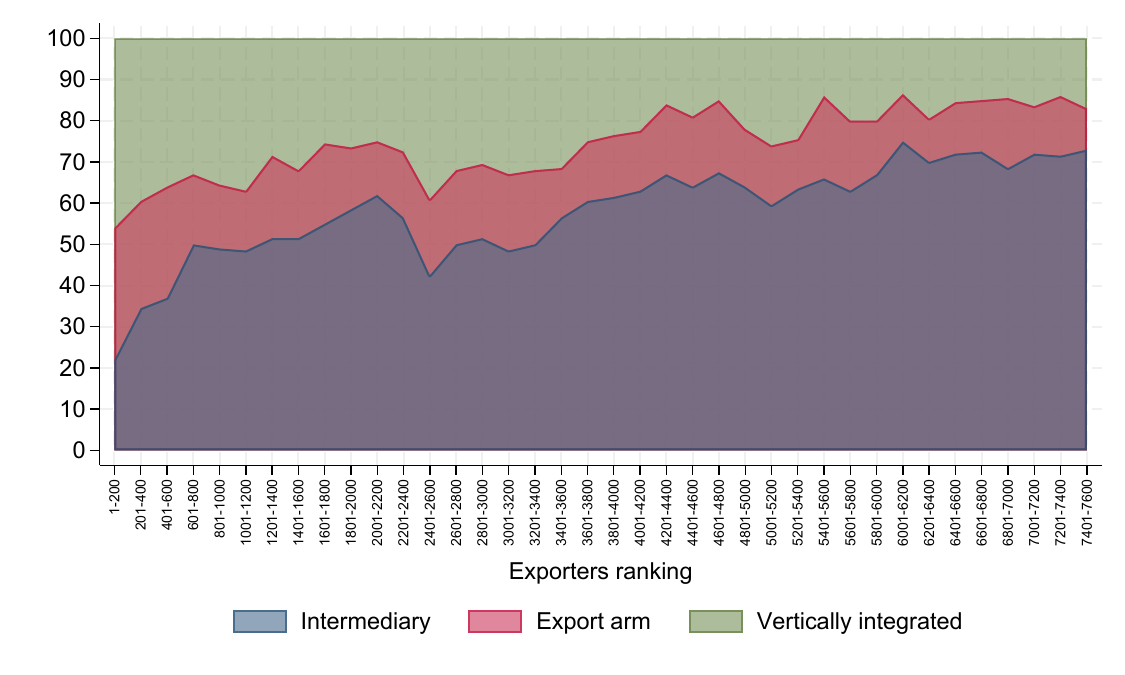}    
    {\footnotesize Source: authors' own elaboration based on AEAT--Customs data merged with Bureau van Dijk's SABI database and LLM classifications.\par}
\end{minipage}
\end{center}
\end{figure}

In sum, focusing on the archetypal intermediary emphasized in trade models, the reclassification reduces intermediaries' share of total exports by 70.3\%. Consequently, a detailed examination of firms classified as wholesalers and retailers points to a much smaller role for pure intermediaries in export activity than the existing literature suggests.

Next, we examine how wholesale and retail exporters differ along key firm and export dimensions. We take intermediaries as the reference group and assess whether export arms and vertically integrated firms differ from them in terms of employment, value added per employee, export value, the number of destination countries, and the number of exported products. We also compare intermediaries with manufacturer exporters.

We estimate the following regression specification:
\begin{equation}
\label{eq:descriptives}
\ln y_{i}=\beta_{1}\text{Export arm}_{i}+\beta_{2}\text{Vertically integrated}_{i}+\beta_{3}\text{Manufacturer}_{i}+\gamma_{s}+\gamma_{k}+\varepsilon_{i}.
\end{equation}

In this equation, $y_{i}$ denotes the outcome of interest for exporter $i$ (e.g., export value). $\text{Export arm}_{i}$ is an indicator equal to one if exporter $i$ is an export arm of a producer firm, $\text{Vertically integrated}_{i}$ is an indicator equal to one if $i$ is a vertically integrated firm, and $\text{Manufacturer}_{i}$ is an indicator equal to one if $i$ is a manufacturer. The terms $\gamma_{s}$ and $\gamma_{k}$ denote employment-decile fixed effects and fixed effects for exporter $i$'s most-exported HS 6-digit product, respectively, and $\varepsilon_{i}$ is an error term. For the employment regression, we omit employment-decile fixed effects.

Table~\ref{tab:premia} reports the regression estimates. Column~1 shows that export arms and vertically integrated firms employ more workers than intermediaries. In particular, employment in export arms and vertically integrated firms is 2.3 and 2.2 times as large, respectively, as in intermediaries. Manufacturers employ more than three times as many workers as pure intermediaries.

\begin{table}[tb]
\begin{center}
\caption{Descriptive regressions}
\label{tab:premia}
{
\def\sym#1{\ifmmode^{#1}\else\(^{#1}\)\fi}
\begin{tabular}{l*{5}{c}}
\toprule
                    &\multicolumn{1}{c}{(1)}&\multicolumn{1}{c}{(2)}&\multicolumn{1}{c}{(3)}&\multicolumn{1}{c}{(4)}&\multicolumn{1}{c}{(5)}\\
                    &\multicolumn{1}{c}{\shortstack{Ln\\Employees}}&\multicolumn{1}{c}{\shortstack{Ln Value added\\per employee}}&\multicolumn{1}{c}{\shortstack{Ln\\Exports}}&\multicolumn{1}{c}{\shortstack{Ln\\Destinations}}&\multicolumn{1}{c}{\shortstack{Ln\\Products}}\\
\midrule
Export arm$_{i}$    &       0.845\sym{a}&       0.278\sym{a}&       0.277\sym{a}&      -0.139\sym{a}&      -0.038       \\
                    &     (0.068)       &     (0.030)       &     (0.096)       &     (0.049)       &     (0.050)       \\
\addlinespace
Vertically integrated$_{i}$&       0.789\sym{a}&       0.052\sym{c}&       0.411\sym{a}&       0.357\sym{a}&      -0.013       \\
                    &     (0.057)       &     (0.027)       &     (0.082)       &     (0.039)       &     (0.038)       \\
\addlinespace
Manufacturer$_{i}$  &       1.140\sym{a}&      -0.070\sym{a}&       0.582\sym{a}&       0.397\sym{a}&      -0.264\sym{a}\\
                    &     (0.051)       &     (0.019)       &     (0.063)       &     (0.030)       &     (0.032)       \\
\midrule
Observations        &       12668       &       12668       &       12668       &       12668       &       12668       \\
Adjusted R$^{2}$         &       0.205       &       0.110       &       0.344       &       0.257       &       0.257       \\
\bottomrule
\end{tabular}
}
\caption*{\begin{footnotesize}Note: The title of each column indicates the dependent variable. Intermediaries is the omitted category. All regressions include fixed effects for the most-exported HS 6-digit product and, except for column~1, employment-decile fixed effects. Standard errors clustered by HS 6-digit product are reported in parentheses. a, b, and c denote statistical significance at the 1\%, 5\%, and 10\% levels, respectively. Data are for 2024.\end{footnotesize}}
\end{center}
\end{table}

Column~2 indicates that, conditional on firm size, value added per employee is higher for export arms and vertically integrated firms than for intermediaries. By contrast, manufacturers exhibit lower value added per employee than intermediaries. Column~3 shows that export arms and vertically integrated firms export more than intermediaries, as do manufacturers. Column~4 shows that intermediaries serve more destinations than export arms, but fewer than vertically integrated firms and manufacturers. Finally, column~5 shows that export arms and vertically integrated firms export the same number of products as intermediaries, whereas manufacturers export fewer products.

Overall, these descriptive regressions reveal sizable differences between intermediaries and both export arms and vertically integrated firms across key firm and export characteristics. Treating all wholesale and retail exporters as a homogeneous group can therefore lead to misleading inferences about their characteristics and export behavior.

\section{Conclusion}
\label{sec:conclusion}

This paper reassesses the role of intermediaries in exports. In both the theoretical and empirical literatures, wholesale and retail firms are typically portrayed as intermediaries: they purchase goods from producers that lack the productivity or trade capabilities to profitably serve foreign markets and use superior trade capabilities to sell those goods abroad. Using Spanish data for 2024, we show that many firms classified as wholesalers or retailers depart markedly from this archetype.

First, many wholesale and retail exporters are owned by producer firms and export exclusively their parent firms' products. Second, some wholesale and retail exporters are vertically integrated, overseeing the full value chain and exporting predominantly their own branded products. Excluding these firms, the number of exporters classified as intermediaries falls by 41.7\%, and their share of total export value falls by 70.3\%. Overall, our findings suggest that intermediaries play a smaller role in exports than previous studies imply.

Our results suggest that, if the Spanish case is representative, policies aimed at strengthening producers’ ability to serve foreign markets may yield larger aggregate export gains than policies focused exclusively on facilitating matches between manufacturers and intermediaries.

\clearpage

\appendix
\numberwithin{equation}{section}

\begin{center}
\textbf{\large Online appendix for\\ ``Reassessing the role of intermediaries in exports''}
\end{center}

\section{Additional tables}
\label{app_sec:additional_tables}

\setcounter{figure}{0}
\renewcommand\thefigure{A.\arabic{figure}}

\setcounter{table}{0}
\renewcommand\thetable{A.\arabic{table}}


\begin{table}[htbp]
\begin{center}
\caption{Classification of wholesale and retail exporters according to each LLM}
\label{tab:detailed_type_ia}
\begin{tabular*}{\textwidth}{@{\extracolsep{\fill}} l r r r r r r }
\toprule
& \multicolumn{2}{c}{ChatGPT} & \multicolumn{2}{c}{Claude} & \multicolumn{2}{c}{Gemini} \\ \cmidrule(lr){2-3} \cmidrule(lr){4-5} \cmidrule(lr){6-7} & Number & \% & Number & \% & Number & \% \\ \midrule
Intermediary&       5,046&        66.4&       3,948&        51.9&       3,791&        49.9\\
Export arm  &       1,218&        16.0&       1,469&        19.3&       1,158&        15.2\\
Vertically integrated&       1,336&        17.6&       2,183&        28.7&       2,651&        34.9\\
\midrule Total&       7,600&       100.0&       7,600&       100.0&       7,600&       100.0\\
\bottomrule
\end{tabular*}
\caption*{\begin{footnotesize}Source: authors' calculations based on LLM classifications.\end{footnotesize}}
\end{center}
\end{table}

\begin{table}[h!tbp]
\begin{center}
\caption{Agreement across LLM classifications (\%)}
\label{tab:detailed_llm_cross}
\begin{tabular*}{\textwidth}{@{\extracolsep{\fill}} l *{3}{>{\raggedleft\arraybackslash}p{3.3cm}} }
\toprule
 & \multicolumn{3}{c}{Second model (\%)} \\ \cmidrule(lr){2-4}
First model & \multicolumn{1}{c}{Intermediary} & \multicolumn{1}{c}{Export arm} & \multicolumn{1}{c}{Vertically integrated} \\ \midrule
\midrule
\textit{ChatGPT} & \multicolumn{3}{c}{\textit{Claude}} \\ \cmidrule[0.05pt](lr){2-4} Intermediary&        69.1&         7.1&        23.8\\
Export arm  &         4.9&        79.7&        15.4\\
Vertically integrated&        30.0&        10.4&        59.6\\
\midrule Overall agreement (\%)&        69.1&            &            \\
\midrule \addlinespace \textit{ChatGPT} & \multicolumn{3}{c}{\textit{Gemini}} \\ \cmidrule[0.05pt](lr){2-4} Intermediary&        70.2&         4.1&        25.7\\
Export arm  &         5.3&        74.4&        20.4\\
Vertically integrated&        14.0&         3.4&        82.6\\
\midrule Overall agreement (\%)&        73.0&            &            \\
\midrule \addlinespace \textit{Claude} & \multicolumn{3}{c}{\textit{Gemini}} \\ \cmidrule[0.05pt](lr){2-4} Intermediary&        73.3&         2.8&        23.9\\
Export arm  &        15.5&        63.8&        20.8\\
Vertically integrated&        30.6&         5.1&        64.3\\
\midrule Overall agreement (\%)&        68.9&            &            \\
\bottomrule
\end{tabular*}
\caption*{\begin{footnotesize}Source: authors' calculations.\end{footnotesize}}
\end{center}
\end{table}



\clearpage

\section{Prompt used to query large language models}
\label{app_sec:prompt}

\setcounter{figure}{0}
\renewcommand\thefigure{B.\arabic{figure}}

\setcounter{table}{0}
\renewcommand\thetable{B.\arabic{table}}

I will provide a list of Spanish firms, including their Tax Identification Number (NIF) and legal name. Your task is to classify each firm into one of the following three groups, based on the criteria below.


\begin{itemize}
    \item \textbf{Type 1: Intermediary.} This group comprises wholesale and retail traders that export goods on behalf of manufacturers with which they have no equity ties. These firms provide an export channel for producers by purchasing their output and managing the export process. Specifically, this category includes:
    \begin{itemize}
        \item multi-brand, multi-supplier distributors
        \item supermarket chains, hypermarkets, large-scale retailers, and independent wholesalers
        \item Spanish subsidiaries and procurement centers linked to independent retail groups
        \item firms classified under 3-digit NACE code 461
    \end{itemize}

    \item \textbf{Type 2: Export arm.} This group comprises Spanish firms that are owned or controlled by a parent company, cooperative, or business group with productive activity in the primary or manufacturing sectors, and that mainly act as commercial, distribution, retail, or export subsidiaries for the products of that parent, cooperative, or group. To be classified as Type 2, the Spanish legal entity itself must \textbf{not} exhibit productive activity in the primary or manufacturing sectors, and must \textbf{not} play the role of the operating parent that centrally coordinates the value chain. This category includes:
    \begin{itemize}
        \item export subsidiaries owned by primary-sector or manufacturing firms that distribute the products of their parent firm or group through specialized commercial units
        \item agricultural cooperatives and agrarian transformation societies that exclusively distribute the output of their member producers, provided that the focal entity itself does not undertake productive activity
        \item producer-owned sales, retail, or distribution subsidiaries that export or commercialize the goods of the parent company or group without undertaking productive activity themselves
        \item Spanish subsidiaries that sell own-brand goods of the parent or group, when the Spanish legal entity itself only commercializes or distributes those goods
    \end{itemize}

    \item \textbf{Type 3: Vertically integrated.} This group comprises Spanish firms that combine wholesale or retail activity with productive activity carried out by the \textbf{same Spanish legal entity}. These firms typically control, within the focal Spanish firm itself, activities such as design, production, processing, manufacturing, or transformation, and export predominantly goods marketed under their own brands.

    A firm should also be classified as \textbf{Type 3} when the focal Spanish legal entity is the \textbf{operating parent or head company} of a business group and there is clear evidence that this legal entity centrally organizes and directs key stages of the value chain for own-brand goods---such as design coordination, sourcing, production planning, procurement, processing, and distribution---even if some manufacturing or processing is executed through controlled subsidiaries or external suppliers. In such cases, Type 3 applies only if the focal Spanish legal entity has an \textbf{active operational role} in coordinating the integrated value chain, rather than being a merely passive holding company.
\end{itemize}

\bigskip

\noindent \textbf{Important legal-entity rule:} classification must be based on the \textbf{specific Spanish legal entity under analysis}, not on the multinational group as a whole. Therefore:

\begin{itemize}
    \item Do \textbf{not} classify a firm as Type 3 solely because its parent company or business group designs, manufactures, or owns the brands of the exported products
    \item Do \textbf{not} classify a firm as Type 3 solely because it sells own-brand goods if the Spanish legal entity itself does not undertake productive activity and does not act as the operating parent that centrally coordinates the value chain
    \item If the parent company or group designs or manufactures the products, but the Spanish subsidiary only commercializes, distributes, retails, or exports them, classify the Spanish firm as \textbf{Type 2}, not Type 3
    \item Do \textbf{not} classify a firm as Type 3 if it is merely a passive holding company, financial holding company, or provider of administrative services, with no evidence of an active operational role in the value chain
\end{itemize}

\bigskip

\noindent \textbf{Priority rule:}
\begin{enumerate}
    \item Assign \textbf{Type 3} if there is clear evidence that the Spanish legal entity itself:
    \begin{itemize}
        \item undertakes productive activity directly, \textbf{or}
        \item acts as the operating parent that centrally coordinates and controls the integrated design, sourcing, production, procurement, and distribution chain for own-brand goods
    \end{itemize}
    \item Otherwise, assign \textbf{Type 2} if the firm is a producer-owned commercial, retail, distribution, or export subsidiary with no productive activity of its own and no operating-parent role
    \item Otherwise, assign \textbf{Type 1}
\end{enumerate}

\bigskip

\noindent \textbf{Interpretation guidance for borderline cases:}
\begin{itemize}
    \item A firm should be classified as \textbf{Type 3} when there is reliable evidence in secondary NACE codes, corporate purpose, annual reports, official company descriptions, or other credible sources that the focal Spanish legal entity itself performs or actively coordinates productive and commercial stages of the value chain
    \item A firm should be classified as \textbf{Type 2} when the focal Spanish legal entity is essentially a sales, retail, distribution, or export vehicle for products designed or manufactured by related firms, even if those products are sold under the group's own brands
    \item In the case of parent companies, distinguish between:
    \begin{itemize}
        \item \textbf{passive parent or holding company}: classify as Type 2 unless there is evidence of direct productive activity
        \item \textbf{operating parent}: classify as Type 3 if there is evidence that the parent company actively coordinates the integrated value chain for own-brand goods
    \end{itemize}
    \item Mere ownership of trademarks, subsidiaries, or brands is \textbf{not} sufficient for Type 3 unless accompanied by evidence of direct productive activity or active operating-parent coordination
    \item If evidence is mixed, ambiguous, or insufficient to establish Type 3, prefer \textbf{Type 2} over Type 3
\end{itemize}

\bigskip

\noindent \textbf{Output instructions:} present the results in a table with the following columns:
\begin{enumerate}
    \item NIF
    \item legal name of the exporting firm
    \item assigned typology (Type 1, Type 2, or Type 3)
    \item concise justification for the classification
\end{enumerate}

\noindent The justification should be brief but specific, and should refer to the role of the \textbf{Spanish legal entity itself} in the value chain. In particular, indicate whether the entity:
\begin{itemize}
    \item acts as an independent intermediary
    \item acts as a producer-owned commercial or export subsidiary without productive activity
    \item undertakes productive activity itself
    \item or acts as an operating parent that centrally coordinates the integrated value chain
\end{itemize}

\bigskip

\noindent When classifying each firm, apply the criteria strictly and conservatively. Assign \textbf{Type 3} only when there is clear evidence of direct productive activity or an active operating-parent role at the level of the focal Spanish legal entity. Otherwise, if the entity is owned by a producer or producer group and mainly commercializes or exports related products without productive activity of its own, assign \textbf{Type 2}. If neither condition applies, assign \textbf{Type 1}.

\end{document}